\documentclass[a4paper,12pt]{article}
\begin{document}

\begin{titlepage}
\thispagestyle{empty}

\begin{flushright}
\end{flushright}
\vspace{5mm}

\begin{center}
{\Large \bf
Superalgebras from $p$-brane actions}
\end{center}

\begin{center}
{\large D. T. Reimers}\\
\vspace{2mm}

\footnotesize{
{\it School of Physics, The University of Western Australia\\
Crawley, W.A. 6009, Australia}
} \\
{\tt  reimers@physics.uwa.edu.au}\\
\end{center}
\vspace{5mm}

\begin{abstract}
\baselineskip=14pt

Two superalgebras associated with $p$-branes are the constraint algebra and the Noether charge algebra. Both contain anomalous terms which modify the standard supertranslation algebra. These anomalous terms have a natural description in terms of double complex cohomology of generalized forms. By retaining fermionic charges and allowing for gauge freedom in the double complex, it is shown that the algebra of conserved charges forms a spectrum with free parameters. The spectrum associated with the Green-Schwarz superstring is shown to contain and generalize the known superalgebras associated with the superstring.

\end{abstract}

\vfill
\end{titlepage}

\renewcommand{\a}[0]{\alpha}
\renewcommand{\b}[0]{\beta}
\newcommand{\g}[0]{\gamma}
\newcommand{\G}[0]{\Gamma}
\renewcommand{\d}[0]{\delta}
\newcommand{\D}[0]{\Delta}
\newcommand{\e}[0]{\epsilon}
\renewcommand{\k}[0]{\kappa}
\newcommand{\m}[0]{\mu}
\newcommand{\n}[0]{\nu}
\newcommand{\p}[0]{\phi}
\renewcommand{\r}[0]{\rho}
\newcommand{\s}[0]{\sigma}
\renewcommand{\t}[0]{\theta}
\newcommand{\x}[0]{\xi}

\newcommand{\sa}[0]{\overrightarrow{\s}}
\newcommand{\nn}[0]{\nonumber}
\newcommand{\gph}[0]{(-g)^{\half}}
\newcommand{\gmh}[0]{(-g)^{-\half}}
\newcommand{\tb}[0]{\overline\theta}
\newcommand{\del}[0]{\partial}
\renewcommand{\bar}[0]{\overline}
\newcommand{\be}[0]{\begin{equation}}
\newcommand{\ee}[0]{\end{equation}}
\newcommand{\bea}[0]{\begin{eqnarray}}
\newcommand{\eea}[0]{\end{eqnarray}}

\newcommand{\half}[0]{\frac{1}{2}}
\newcommand{\third}[0]{\frac{1}{3}}
\newcommand{\sixth}[0]{\frac{1}{6}}
\newcommand{\dtgtl}[1]{d\overline{\theta}\Gamma_{#1}\theta}
\newcommand{\dtgtu}[1]{d\overline{\theta}\Gamma^{#1}\theta}
\newcommand{\tgdotl}[1]{\overline{\theta}\Gamma_{#1}\del_{1}\theta}
\newcommand{\tgdotu}[1]{\overline{\theta}\Gamma^{#1}\del_{1}\theta}
\newcommand{\gtl}[2]{(\Gamma_{#1} \theta)_{#2}}
\newcommand{\gtu}[2]{(\Gamma^{#1} \theta)_{#2}}

\newtheorem{theorem}{Theorem}

\section{Introduction}

The action for a $p$-brane is comprised of the kinetic term and the WZ (Wess-Zumino) term \cite{green84,Bergshoeff87,ach87}. The WZ term ensures that a local ``$\k$ symmetry" is present which means that only half the fermionic degrees of freedom are physical \cite{siegel83-kappa}. The Lagrangian is not manifestly invariant under the global action of the supertranslation group (it is not ``left invariant") due to quasi-invariance (invariance up to a total derivative) of the WZ term. The WZ term is the pullback of a ($p+1$)-form $B$ which is a potential for a field strength $H$. Although $H$ is left invariant, in standard superspace it is impossible to find a left invariant potential $B$. In terms of CE (Chevalley-Eilenberg) cohomology \cite{chevalley48} this means that $H$ is a nontrivial cocycle. In fact, $H$ is characterized as the unique nontrivial CE $(p+2)$-cocycle of dimension $p+1$ \cite{azc89-2}. There are two avenues of research that have resulted from this fact.

The first area of research concerns topological charge algebras. The Noether charges associated with left invariance of an action are phase space generators of the left group action (``left generators''). For manifestly left invariant actions, the algebra of Noether charges is the same as the underlying algebra of symmetries. This is the ``minimal algebra'' of Noether charges. However, Lagrangians are often quasi-invariant under the action of symmetry transformations. In this case the Noether charges need to be modified in order to ensure their conservation. The conserved charges then obey an algebra which is a modification of the minimal algebra by a topological ``anomalous term'' \cite{witten78}. This is the case for p-brane actions, where quasi-invariance of the WZ term under the action of the supertranslation group means that the Noether charges satisfy an algebra which is an extension of the supertranslation algebra by a topological term \cite{azc89}. In the conventional formulation of superspace, the fermionic directions have trivial topology \cite{dewitt}. In this case, the anomalous term simplifies to a form which can be related to PBRS (partial breaking of rigid supersymmetry) \cite{sorokin97,townsend97}. The algebra of constraints for the action is also modified in the presence of the WZ term \cite{azc91}. The constraints can be identified with generators of the right group action (``right generators"), thus leading to a modified algebra of right generators. The modified algebras of Noether charges and constraints can also be related to a construction involving ghost fields  \cite{azc91}. A BRST style ``ghost differential" $s$ acting on an infinite dimensional ``loop superspace" is introduced. The anomalous term is then the result of solving cohomological descent equations. A similar construction of a finite dimensional nature has also been considered \cite{Bergshoeff98}.

In a second line of research, superspaces associated with extensions of the supertranslation algebra have been discovered which allow manifestly left invariant WZ terms to be constructed for the $p$-brane action \cite{siegel94,bergshoeff95,chrys99}. The resulting actions can be considered equivalent to the standard action since the Lagrangians differ only by a total derivative (and the extra superspace coordinates appear only in this derivative). Due to manifest left invariance of the Lagrangian, the Noether charges are not modified and they satisfy the minimal algebra (which in this case reflects the underlying extended supertranslation algebra). There is partial correspondence with the standard superspace formulation of the action here. The anomalous term in the standard superspace formulation can be identified with a corresponding term in the minimal algebra of an extended superspace formulation \cite{chrys99}. However, the full extended superalgebra is not generated in this way. In the extended superspace formulation there are fermionic Noether charges which complete the full algebra. However, in the standard superspace formulation, the assumption of trivial fermionic topology prevents the existence of any analog of these fermionic charges. As a result, one obtains only the ``first line" of the full superalgebra. In this paper we will show that a full agreement between the algebras of the different action formulations is possible. In doing so we will also address the fact that there is more than one extended superspace which allows a manifestly left invariant WZ term to be constructed. Which of these extended superspaces should be generated by the anomalous term of the standard action?

There are hints that incorporating fermionic charges (whether they are topological in nature or arise from fermionic boundary conditions) in brane theory may yield interesting results. For example, the action of supersymmetries on bosonic charges clearly produces fermionic charges \cite{hatsuda00}. Should these charges vanish? Quantizing in the standard flat background allows one to choose a trivial representation for the fermionic charges. However, in certain superspaces nonvanishing fermionic charges are actually \textit{required} \cite{peeters03}. The construction of topological anomalous terms has always allowed for nontrivial topology of the bosonic coordinates (otherwise even the classical charges would vanish). However, the terms that would result from nontrivial fermionic topology have usually been omitted.

In this paper we investigate $p$-brane superalgebras by focussing on the underlying double complex cohomology of the anomalous terms. A number of new results follow. The anomalous terms of the algebras of left/right generators are shown to derive from representatives of a single complex cocycle associated with the $p$-brane. The presence of gauge freedom for these representatives leads to the identification of a new freedom in the anomalous term of the Noether charge algebra. It follows that this anomalous term is not well defined as a form, but as an entire \textit{cohomology class} $[M]$. In the standard superspace background, $[M]$ is shown to be a unique, nontrivial class which may be constructed on the basis of the same dimensionality and Lorentz invariance requirements used to construct $H$ in \cite{azc89-2}. It is also shown that $[M]$ defines a spectrum of extended superalgebras. When fermionic charges are allowed, these superalgebras are realized as the topological charge algebras of the action.

The construction is then applied to the GS (Green-Schwarz) superstring. The topological charges are identified as extra generators of the Noether charge algebra. The resulting topological charge algebra is shown to be a one parameter spectrum of extended superalgebras. When fermionic charges are retained, this spectrum contains three extended algebras of interest. The first is an algebra developed by Green, which has a fermionic ``central" extension \cite{green89}. The second is an algebra which extends the Green algebra by a noncentral bosonic generator. Both of these algebras have been used to construct string Lagrangians that have \textit{manifest} left invariance, and are thus of physical significance \cite{siegel94,bergshoeff95,chrys99}. The third algebra, which is of the type considered in \cite{hatsuda00,peeters03}, results from the action of supersymmetry on the bosonic charge. It thus emerges naturally that if fermionic charges are retained, \textit{all} the known extended algebras of the superstring appear in the spectrum of topological charge algebras of the standard action. Since the spectrum is not simply obtained by rescaling known algebras, new superalgebras also result from the process.

The structure of this paper is as follows. In section \ref{sec:Extended algebras} our conventions are outlined and the properties of $p$-branes are reviewed. The extended superspaces used for the GS superstring are also presented. In section \ref{The p-brane double complex} the ghost differential $s$ is defined, and is then used to define a superspace double complex. An exactness theorem for $s$ is presented, and a total differential $D$ is defined which is shown to evaluate the CE cohomology of the WZ term. It is shown that the $p$-brane has a naturally associated $D$ cocycle which is defined by the $(p+2)$-form $H$. In section \ref{sec:Algebra modifications} the construction of topological anomalous terms is reviewed in a fully integrated approach. It is shown that the anomalous term defines an extension of the underlying superalgebra by an ideal. Modified generators of the right action are defined. The resulting modified algebra is shown to derive from the representative $H$ of the $p$-brane $D$ cocycle. The relationship between the right generator algebra and the constraint algebra is given. Cohomological properties of the anomalous term that follow from the CE properties of $H$ are presented in two theorems. The first theorem defines the anomalous term as a cohomology class. The second theorem states that in standard superspace this class can be constructed using uniqueness of the cocycle and dimensional analysis. In section \ref{sec:Application to the GS superstring} the construction is applied to the GS superstring. Both standard and extended superspace actions are investigated. We point out that the reader may find it helpful to read the explicit examples of this section in conjunction with the general theory of the preceding sections. The algebra of right generators and the constraint algebra are evaluated. Both are shown to agree with the cocycle construction. The topological charge algebra is found by solving the descent equations. The most general gauge transformation of the anomalous term is shown to contain a single degree of freedom. This freedom is used to generate a spectrum of algebras that includes the known extended superalgebras of the superstring. Properties of the extended superspace actions are shown to be consistent with the general construction. In section \ref{sec:Conclusion} some comments on future directions for research are made.

\section{Preliminaries}
\label{sec:Extended algebras}

\subsection{$p$-branes}
\label{sec:Preliminaries}

The superalgebra of the supertranslation group is\footnote{The charge conjugation matrix will not be explicitly shown. It will only be used to raise/lower indices on gamma matrices, which have the standard position $\Gamma^{\a}{}_{\b}$. $\Gamma_{\a\b}$ is assumed to be symmetric. Majorana spinors are assumed throughout (thus, for example, $\bar\t_{\a}=\t^{\b}C_{\b\a}$).}:
\bea
    \{Q_{\alpha},Q_{\beta}\}&=&\Gamma^{a}{}_{\alpha\beta}P_{a}.
\eea
The corresponding group manifold can be parameterized:
\be
    g(Z)=e^{x^{a}P_{a}}e^{\theta^{\a}Q_{\a}},
\ee
where $Z$ is the combined notation for coordinates:
\[
    Z^{A}=\{x^{a},\theta^{\a}\}.
\]
This group can be constructed as the coset space consisting of the super-Poincar\'{e} algebra modulo the Lorentz subgroup, however for our purposes this is an unnecessary complication. In this paper it is valid to assume that expressions are Lorentz invariant if upper indices are contracted with lower ones.

The left vielbein is defined by:
\begin{eqnarray}
    L(Z)&=&g^{-1}(Z)dg(Z)\\
    &=&dZ^{M}L_{M}{}^{A}(Z)T_{A}\nn,
\end{eqnarray}
where $T_A$ represents the full set of superalgebra generators. The right vielbein is defined similarly:
\begin{eqnarray}
    R(Z)&=&dg(Z)g^{-1}(Z)\\
    &=&dZ^{M}R_{M}{}^{A}(Z)T_{A}\nn.
\end{eqnarray}
The left group action is defined by:
\be
	g(Z')=g(\e)g(Z),
\ee
where $\e^A$ is an infinitesimal constant. The corresponding superspace transformation is generated by the operators:
\be
	\label{2:scalar left gen}
	 Q_{A}=R_{A}{}^{M}\del_{M},
\ee
where $R_{A}{}^{M}$ are the inverse right vielbein components, defined by:
\be
    R_{A}{}^{M}R_{M}{}^{B}=\d_{A}{}^{B}.
\ee
$Q_A$ are generators of the left group action, and will be referred to as the ``left generators." Forms that are invariant under the global left group action will be called ``left invariant." The vielbein components $L^{A}$ are left invariant by construction. Their explicit structure is:
\begin{eqnarray}
    L^{a}&=&dx^{a}-\half \dtgtu{a}\\
    L^{\alpha}&=&d\theta^{\alpha}\nn.
\end{eqnarray}
Indices $A,B,C,D$ will be used to indicate components with respect to this basis, while $M,N,L,P$ will be used for the coordinate basis. The right group action is defined by:
\be
	g(Z')=g(Z)g(\e).
\ee
The corresponding superspace transformation is generated by the operators:
\be
	\label{right generators}
D_{A}=L_{A}{}^{M}\del_{M},
\ee
where $L_{A}{}^{M}$ are the inverse left vielbein components, defined by:
\be
    L_{A}{}^{M}L_{M}{}^{B}=\d_{A}{}^{B}.
\ee
$D_A$ are generators of the right supertranslation group action, and will be referred to as the ``right generators." They are also commonly known as ``supercovariant derivatives" since they commute with the left generators as a result of the associativity of group multiplication. However, unlike the $Q_{A}$ they do not generate global symmetries of the action. The left and right vielbein and inverse vielbein components have been evaluated and placed in appendix \ref{sec:app:standard vielbein components} for reference.

The NG (Nambu-Goto) action for a $p+1$ dimensional manifold embedded in the background superspace is:
\be
	S=-\int d^{p+1}\s\sqrt{-g}.
\ee
The integral is over the $p+1$ dimensional ``worldvolume," which has coordinates $\s^{i}$ and is embedded in superspace. The worldvolume metric $g_{ij}$ is defined using pullbacks of the left vielbein:
\bea
    L_{i}{}^{A}&=&\del_{i}Z^{M}L_{M}{}^{A}\\
    g_{ij}&=&L_{i}{}^{a}L_{j}{}^{b}\eta_{ab}\nn,
\eea
and $g$ denotes $\det g_{ij}$. A $p$-brane is the $\k$-symmetric generalization of the NG action. The $p$-brane action is:
\be
	\label{3:p-brane action}
	S=-\int d^{p+1}\s\sqrt{-g}+\int B.
\ee
The first term of the action is the ``kinetic" term. The second term is the WZ term, which is the integral over the worldvolume of the pullback of a superspace form $B$. $B$ is defined by the property\footnote{Wedge product multiplication of forms is understood.}:
\bea
    \label{3:H def}
    dB&=&H\\
    &\propto& d\t^{\a}d\t^{\b}L^{a_{1}}\ldots L^{a_{p}}(\G_{a_{1}\ldots a_{p}})_{\a\b}.\nn
\eea
The proportionality constant depends on $p$ and is determined by requiring $\k$ symmetry of the action. There are certain identities required to ensure the consistency of this definition. Firstly, closure of $H$ requires a Fierz identity:
\be
    \label{3:p-brane fierz}
    \G^{[a_{1}...a_{p}]}{}_{(\a\b}\G_{a_p\d\e)}=0.
\ee
This condition on the gamma matrices can only be satisfied for certain combinations of $p$ (spatial dimension of the brane) and $d$ (superspace dimension)  \cite{evans88}. The allowed values of $(p,d)$ (called the ``minimal branescan") are such that:
\be
    \label{3:p brane Gamma symmetry requirement}
    (\G_{[a_{1}...a_{p}]})_{\a\b}=(\G_{[a_{1}...a_{p}]})_{\b\a}.
\ee
This ensures that $H$ can be nonzero.

\subsection{Green algebra}

The $p$-brane action (\ref{3:p-brane action}) can also be used in extended superspace backgrounds. In the general construction we will not specify the background being used in order to allow for this possibility\footnote{However, extended backgrounds that are extensions of standard superspace by an ideal are most naturally applied (see appendix \ref{sec:app:algebra conditions}). The extended algebras associated with $p$-branes have this property.}. In section \ref{sec:Application to the GS superstring} we will consider the GS superstring in both standard and extended superspaces. There are two known extended superspaces which allow the construction of manifestly left invariant superstring WZ terms. The first is described by a superalgebra that was introduced by Green \cite{green89}. It has a fermionic generator $\Sigma^\a$ that defines a central extension of the supertranslation group\footnote{Of course, when Lorentz generators are included, $\Sigma^\a$ is no longer central.}:
\bea
    \label{2:extalg}
    \{Q_{\alpha},Q_{\beta}\}&=&\Gamma^{a}{}_{\alpha\beta}P_{a}\\
    \ [Q_{\b},P_{a}]&=&\Gamma_{a\b\g}\Sigma^{\g}\nn.
\eea
The corresponding group manifold can be parameterized\footnote{Parameterizations are not unique. In particular we note the Green algebra can alternatively be parameterized to yield a linear realization of the left group action \cite{derig97}.}:
\begin{eqnarray}
    g(Z)=e^{x^{a}P_{a}}e^{\theta^{\a}Q_{\a}}e^{\phi_{\b}\Sigma^{\b}},
\end{eqnarray}
where:
\[
    Z^{A}=\{x^{a},\theta^{\a},\phi_{\a}\}.
\]
Standard superspace is obtained by omitting the extra generator $\Sigma^{\a}$ (and its associated coordinate $\p_{\a}$). The resulting left vielbein components are:
\begin{eqnarray}
    L^{a}&=&dx^{a}-\half \dtgtu{a}\\
    L^{\alpha}&=&d\theta^{\alpha}\nn\\
    L_{\alpha}&=&d\phi_{\alpha}-dx^{b}\gtl{b}{\alpha}+\sixth\dtgtu{b}\gtl{b}{\alpha}\nn.
\end{eqnarray}
The left and right vielbein and inverse vielbein components for the Green algebra have been evaluated and placed in appendix \ref{sec:app:green algebra vielbein components}.

\subsection{Extended Green algebra}

Addition to the Green algebra of a noncentral bosonic generator $\Sigma^{a}$ results in the extended Green algebra \cite{bergshoeff95, chrys99}:
\bea
    \label{2:ext Green algebra}
    \{Q_{\alpha},Q_{\beta}\}&=&\Gamma^{a}{}_{\alpha\beta}P_{a}+\Gamma_{a}{}_{\alpha\beta}\Sigma^{a}\\
    \ [Q_{\b},P_{a}]&=&\Gamma_{a\b\g}\Sigma^{\g}\nn\\
    \ [Q_{\b},\Sigma^{a}]&=&\Gamma^a{}_{\b\g}\Sigma^{\g}\nn.
\eea
The Green algebra results from the reduction:
\bea
    P'_a&=&P_{a}+\eta_{ab}\Sigma^{b}\\
    \Sigma'^{\a}&=&2\Sigma^{\a},\nn
\eea
where $\eta_{ab}$ is the Minkowski metric. The extended Green algebra group manifold can be parameterized:
\be
    g(Z)=e^{x^{a}P_{a}}e^{y_{b}\Sigma^{b}}e^{\theta^{\a}Q_{\a}}e^{\phi_{\b}\Sigma^{\b}},
\ee
with coordinates\footnote{Coordinate indices will not be raised/lowered in this paper. In the notation being used $\{Z^{a},Z^{\a},Z_{a},Z_{\a}\}$ are all independent coordinates.}:
\[
    Z^{A}=(x^{a},\theta^{\a},y_{a},\phi_{\a}).
\]
The left vielbein components are found to be:
\begin{eqnarray}
    L^{a}&=&dx^{a}-\half \dtgtu{a}\\
    L^{\alpha}&=&d\theta^{\alpha}\nn\\
    L_{a}&=&dy_{a}-\half \dtgtl{a}\nn\\
    L_{\alpha}&=&d\phi_{\alpha}-dx^{b}\gtl{b}{\alpha}-dy_b\gtu{b}{\alpha}+\third\dtgtu{b}\gtl{b}{\alpha}\nn.
\end{eqnarray}
The left/right vielbein and inverse vielbein components for the extended Green algebra have been evaluated and placed in appendix \ref{sec:app:ext green algebra vielbein components}.

\section{Double complex for the $p$-brane}
\label{The p-brane double complex}

\subsection{Cocycles from WZ terms}
\label{sec:WZ terms as a double complex}

The exterior derivative $d$ together with the space of differential forms constitutes the de Rham complex. The operator $d$ is nilpotent (i.e. $d^{2}=0$) and can therefore be used to define cohomology classes. The $n$-th de Rham cohomology is the set of equivalence classes:
\be
    H_d^{n}=Z^{n}/B^{n}
\ee
where $Z^{n}$ are the closed $n$-forms (i.e. those in the kernel of $d$) and $B^{n}$ are the exact $n$-forms (those in the image of $d$). The de Rham complex can be extended into a double complex by the addition of a second nilpotent operator that commutes with $d$. The operator used in this paper is a ``ghost differential" $s$. This operator was introduced in \cite{azc91} acting on an infinite dimensional ``loop superspace." We now define the analogous operator for use on finite dimensional superspaces. The introduction of a ghost partner $e^{A}$ for each coordinate is required. The ghost fields have the opposite grading to coordinates:
\bea
    \ [e^{A},Z^{M}\}&=&0\\
    \{e^{A},e^{B}]&=&0\nn,
\eea
where $[\quad,\quad\}$ and $\{\quad,\quad]$ are the graded commutator/anticommutator:
\bea
    [X_A,X_B\}&=&-(-1)^{AB}[X_B,X_A\}\\
    \{X_A,X_B]&=&(-1)^{AB}\{X_B,X_A]\nn.
\eea
They are independent of the fields $Z^{M}$, and hence satisfy $de^{A}=0$. A general element of the double complex is a ``ghost form valued differential form." The space of all such ``generalized forms" of differential degree $m$ and ghost degree $n$ will be denoted by $\Omega^{m,n}$. The collection of these spaces will be denoted $\Omega^{*,*}$. Generalized forms $Y\in \Omega^{m,n}$ will be written using a comma to separate ghost indices from space indices:
\be
    Y=e^{B_{n}}\ldots e^{B_{1}}L^{A_{m}}\ldots L^{A_{1}}Y_{A_{1}\ldots A_{m},B_{1}\ldots B_{n}}\frac{1}{m!n!}.
\ee
We then define the ghost differential by the following properties:
\begin{itemize}
\item
$s$ is a right derivation\footnote{Our conventions are such that $d$ is also a right derivation (with respect to the differential degree).}. That is, if $X$ and $Y$ are generalized forms and $n$ is the ghost degree of $Y$ then:
\be
    s(XY)=Xs(Y)+(-1)^{n}s(X)Y.
\ee
\item
If $X$ has ghost degree zero then:
\be
    \label{3:s operator Q part}
    sX=e^{A}Q_{A}X.
\ee
$Q_{A}$ denotes a Lie derivative with respect to the vector field (\ref{2:scalar left gen}) associated with the global left action.
\item
\be
    se^{A}=\half e^{C}e^{B}t_{BC}{}^{A},
\ee
where $t_{BC}{}^A$ are the structure constants of the superalgebra associated with the background superspace (we henceforth refer to this superalgebra as the ``background superalgebra").
\end{itemize}
One verifies that\footnote{To prove nilpotency of $s$ one needs to use the Jacobi identity for the background superalgebra.}:
\bea
    \label{6:sdprops}
    s^{2}&=&0\\
    \ [s,d]&=&0\nn.
\eea
Hence $s$ extends the de Rham complex into a double complex. $s$ is similar to a BRST operator in that it requires the introduction of ghost fields; however unlike a BRST operator it has not been derived from constraints or gauge symmetries.

There is a total differential $D$ that is naturally associated with the double complex:
\bea
    \label{3:d-s differential D}
    D&=&s+(-1)^{n+1}d\\
    D^{2}&=&0\nn,
\eea
where $n$ is the ghost degree of the generalized form upon which $D$ acts. The spaces $\Omega^l_{D}$ of the single complex upon which $D$ acts are the sum along the anti-diagonal of the spaces of the double complex:
\begin{figure}[t]
\begin{center}
\begin{picture}(120,130)(0,-10)
\put(0,18){\vector(1,0){123}}
\put(0,18){\vector(0,1){85}}
\put(10,10){\makebox(50,50){
\large $\begin{array}{cccccccc}
& 3\ & dB & & &\\
& 2\ & B &\diamondsuit& &\\
\uparrow & 1\ & & W &\diamondsuit&\\
d & 0\ & & & N & sN\\
& & 0 & 1 & 2 & 3\\
& & s & \rightarrow & &
\end{array}$
}}
\end{picture}
\caption{Descending sequence for the string}
\label{6fig:basic HB box}
\end{center}
\end{figure}
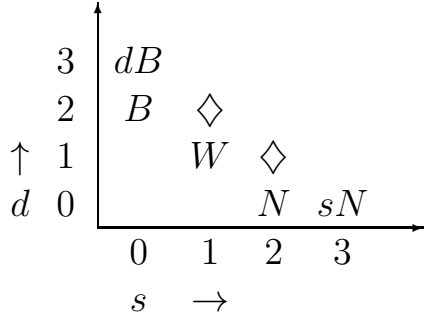
\be
	\Omega_{D}^{l}=\{\oplus\Omega^{m,n}:\qquad m+n=l\}.
\ee
The $l$-th cohomology of $D$ is:
\be
	H_D^{l}=Z_D^{l}/B_D^{l},
\ee
where $Z_D^{l}$ are the $D$ closed generalized $l$-forms (``$D$ cocycles"), and $B_D^{l}$ are the generalized $l$-forms in the image of $D$ (``$D$ coboundaries"). The restriction of $H_D^l$ to representatives within $\Omega^{m,l-m}$ will be denoted $H^{m,l-m}$. The representatives of $H^{m,0}$ can be used to define descending cohomology sequences. We now illustrate this for the $(p+2)$-form $H$ that defines the WZ term.

Firstly, $H$ is a left invariant, closed form with ghost number zero. It is therefore closed under both $s$ and $d$. Using the fact that $s$ and $d$ commute, $sH=0$ implies that $dsB=0$, and thus $sB=-dW$ for some $W\in \Omega^{p,1}$. This argument does not apply globally, but is valid on every coordinate patch\footnote{The fields of the double complex can be viewed as Cech cochains. In this case an expression like $dB$ represents something closed but not necessarily exact. The de Rham triviality of such fields does not affect the double complex cohomology studied in this paper.}. The same logic that was applied to $B$ can then be applied to $W$. This gives $sW=dN$ for some $N\in \Omega^{p-1,2}$. For the string, the last nonzero element of the sequence is $sN\in H^{0,3}$. For a $p$-brane, the sequence continues until we reach an element of $H^{0,p+2}$. The descending cohomology sequence can be graphically depicted using a ``tic-tac-toe box" \cite{bott82}. The string case is depicted in figure \ref{6fig:basic HB box}. The symbol $\diamondsuit$ indicates ``zero with respect to the operator $D$." Precisely, for a $p$-brane, denote the ``potentials" of the sequence by $B^{p+1-m,m}$ (e.g. $W=B^{p,1}$). Then each $\diamondsuit$ represents a relation:
\be
    sB^{p-m+1,m}+(-1)^{m}dB^{p-m,m+1}=0.
\ee
These are the ``descent equations" (note that the first descent equation, not represented in the above, is $H=dB^{p+1,0}$).

We have defined the tic-tac-toe construction on the double complex so that its endpoints would be linked via a coboundary of the $D$ complex. For example, in the string case:
\be
	-dB\oplus sN=D(B\oplus W\oplus N).
\ee
That is, $H=dB\in H^{3,0}$ is $D$ cohomologous to $sN\in H^{0,3}$. We may write this as:
\[H\simeq sN.\]
In general one finds that:
\be
    H\simeq sB^{p+1-m,m} \qquad\forall m.
\ee
The $D$ cocycle represented by $H$ can therefore be alternatively represented by $s$ acting on any of the potentials of the sequence.

We will call a nilpotent operator ``exact" if its associated cohomology is trivial. For example, the de Rham differential $d$ on an open set is exact; the cohomology $H_d$ is trivial as a result of the Poincar\'{e} lemma. That is, given $Y\in H_{d}^{m}$, then for all $m\geq 1$ we can write $Y=dX$ for some $X\in \Omega_{d}^{m-1}$. Note that the ``exactness" of an operator is dependent on the space upon which it acts. By definition, $d$ is not exact (globally) on a manifold that possesses nontrivial de Rham cohomology. There are important consequences for $D$ cohomology if we can show that the ghost differential $s$ is exact.
\begin{theorem}[exactness]
$s$ is exact on open sets.
\end{theorem}

To prove this we find a chain map for which the operators $s$ and $d$ become ``dual" to each other. A chain map between two complexes is one that commutes with the differentials of the complexes. In our case, the required chain map $\Psi$ must satisfy:
\bea
    \Psi(d)\Psi(Y)&=&\Psi(dY)\\
    \Psi(s)\Psi(Y)&=&\Psi(sY)\nn
\eea
for any $Y\in \Omega^{*,*}$. The chain map is the ``check map" defined by:
\bea
    \Psi:\Omega^{*,*}&\rightarrow&\check\Omega^{*,*}\\
    L^{A}&\rightarrow&e^{A}\nn\\
    e^{A}&\rightarrow&R^{A}\nn.
\eea
The map takes $(m,n)$-forms to $(n,m)$-forms. On $\check\Omega^{*,*}$ we have the operators $\check s$ and $\check d$ defined by:
\begin{itemize}
\item
\be
    \check s=d.
\ee
\item
$\check d$ is a right derivation.
\item
If $X$ has ghost degree zero then:
\be
    \check dX=e^{A}D_{A}X,
\ee
where $D_{A}$ is a Lie derivative with respect to the vector field (\ref{right generators}) associated with the global right action.
\item
\be
    \check de^{A}=-\half e^{C}e^{B}t_{BC}{}^{A}.
\ee
\end{itemize}
If we think of $s$ as a generalized left variation, then $\check d$ is the analogous right variation. The check map is clearly invertible. Let $Y$ be any $s$ closed generalized form of ghost degree one or more over an open set. Then, using $\check s=d$ and the exactness of $d$ on an open set, one shows that $Y$ is an $s$ coboundary:
\bea
    sY&=&0\\
    \Rightarrow \check s\check Y&=&0\nn\\
    \Rightarrow \check Y&=&\check s\check X\nn\\
    \Rightarrow Y&=&sX\nn.
\eea
Therefore $s$ is exact on open sets since we have $H^m_{s}=H^m_{d}$.

In \cite{azc89-2} it was shown that CE cohomology can be restated as the restriction of de Rham cohomology to left invariant forms. Now, the $(p+2)$-form $H$ is a $D$ coboundary when it can be written $H=DB$. Equivalently, $H$ is a $D$ coboundary if a left invariant potential $B$ can be found. This is precisely the definition of a trivial CE cocycle. A nontrivial $D$ cocycle is one for which we must necessarily have $sB\neq 0$, which is equivalent to the definition of a nontrivial CE cocycle. CE cohomology is therefore the restriction of $D$ cohomology to forms that have ghost degree zero. $H_D$ is the natural extension of CE cohomology into the double complex $\Omega^{*,*}$. Since $s$ is exact, we may reverse descending tic-tac-toe sequences into ascending ones, starting with any element of $H_D$ and finding an associated left invariant element of $H_d$. This establishes an isomorphism between $H_D$ and CE cohomology that would not exist if $s$ were not exact.

\subsection{Gauge freedom}

Using the tic-tac-toe construction, the form $H\in H^{p+2,0}$ may be identified with any of the other representatives $sB^{p+1-m,m}$ of the $p$-brane $D$ cocycle. This is a well defined map between $H^{m,n}$ cohomologies, but \textit{not} between the forms themselves. In general there is gauge freedom for representatives. Although this freedom is associated with $D$ coboundaries, there is no reason for these coboundaries to be exact. In this way we will see that the gauge freedom can affect the topological charge algebra.

We now explicitly derive the gauge transformations for the string. Consider the relation $H=dB$. Given $H$, this defines $B$ only up to a closed form. Thus, given a solution $B$, the alternative solution $B'=B-d\psi$ is equally valid. We write this as:
\be
    \label{delta B=d psi}
    \D B=-d\psi.
\ee
In an extended superspace that allows a manifestly invariant WZ term, a transformation of this type is all that separates the standard WZ term from the invariant one (see section \ref{sec:Application to the GS superstring} for an explicit example). What then is the effect of the transformation (\ref{delta B=d psi}) on $W$? Since the variation $\D$ commutes with $s$ and $d$, we have:
\bea
    d\D W&=&\D dW\\
    &=&-\D sB\nn\\
    &=&ds\psi\nn.
\eea
The general solution may be written:
\be
    \label{3:delta W}
    \D W=s\psi+d\lambda,
\ee
where $\lambda$ is a new gauge field. The gauge transformations of the field $N$ are derived similarly. Directly from (\ref{3:delta W}) we have:
\bea
    d\D N&=&\D dN\\
    &=&\D sW\nn\\
    &=&ds\lambda\nn.
\eea
This has the general solution:
\be
    \D N=s\lambda+C,
\ee
where $C$ is a $d$ closed $(0,2)$-form. If one progressed in the other direction (an ascending sequence starting from $sN$) one would also find an $s$ closed $(2,0)$-form gauge field $C'$ for $B$. One can then write the gauge transformations in totality as:
\be
    \D(B\oplus W\oplus N)=D(\psi\oplus \lambda)\oplus C \oplus C'.
\ee
One verifies that each potential has two gauge fields: one that is $d$ closed and one that is $s$ closed.
These gauge transformations are additive. For example, the field $W$ has two gauge transformations: one for $\psi$ and one for $\lambda$, with $\D W$ given by (\ref{3:delta W}). The gauge fields are independent (they are not required to satisfy descent equations like those that relate $B$, $W$ and $N$). They may also affect more than one field. For example, $\psi$ is a transformation that leaves $dB$ invariant ($\D B=-d\psi$), and also $sW$ invariant ($\D W=s\psi$). Although $sB$ and $dW$ are not gauge invariant, the $\psi$ transformation is such that the descent equation $sB=-dW$ is true in all gauges. The construction ensures that in general, descent equations are preserved by the gauge transformations. The gauge transformations are identically the same as the $D$ coboundaries as a result of the exactness of the operators $s$ and $d$. The alternative representatives $sB$, $sW$ and $sN$ of the $D$ cocycle defined by $H$ are therefore well defined elements of their $H^{m,n}$ cohomologies.

\section{Algebra modifications}
\label{sec:Algebra modifications}

\subsection{The algebra of left generators}
\label{sec:The algebra of left generators}

The action is formulated in terms of $(Z^{M},\dot Z^{M})$, which may be viewed as coordinates for the superspace tangent bundle. The Hamiltonian formulation of dynamics is cast in terms of coordinates $Z^{M}$ and their associated conjugate momenta $P_{M}$, which together constitute the ``phase space." The momenta are defined by:
\be
    \label{3:momentadef}
    P_{M}=\frac{\del L}{\del \dot Z^{M}}.
\ee
The phase space can be viewed as coordinates for the superspace cotangent bundle. The Lagrangian then provides a map (a Legendre transform), defined by (\ref{3:momentadef}), from the tangent bundle to the cotangent bundle.

We use the following fundamental (graded) Poisson brackets on phase space\footnote{Brackets of unspecified type will in general be Poisson brackets. Exceptions should be clear within context.}:
\be
    \label{3:basicbrackets}
    \ [P_{M}(\s),Z^{N}(\s')\}=\d_{M}{}^{N}\d(\sa-\sa'),
\ee
where it is assumed $\s'^{0}=\s^{0}$ (i.e. equal time brackets). The Dirac delta function notation is shorthand for the product of the $p$ delta functions on the spatial coordinates of the worldvolume. One can use (\ref{3:basicbrackets}) and the following Poisson bracket identities to evaluate general brackets:
\bea
    \label{3:bracket identities}
    \ [X_{A},X_{B}X_{C}\}&=&[X_{A},X_{B}\}X_{C}+(-1)^{AB}X_{B}[X_{A},X_{C}\}\\
    \ [X_{A},X_{B}(Y)\}&=&[X_{A},Y^{N}\}\frac{\del X_{B}}{\del Y^{N}}\nn.
\eea
The above relations can all be derived from an integral form of the Poisson bracket, which can be useful for certain proofs. The form we use is:
\be
    [X_{A},X_{B}\}=\int d^{p}\s\frac{\d X_{A}}{\d P_{M}(\s)}\frac{\d X_{B}}{\d      Z^{M}(\s)}(-1)^{MA+M}-(-1)^{AB}[A\leftrightarrow B].
\ee
We define the following regularly used ``bar map" by its action on superspace forms:
\be
	\label{3:form correspondence}
	\bar Y^{m-p,n}(\s)=(-1)^{p(p+m+1)}i_{\del_{1}}.\ .\ .\ i_{\del_{p}}Y^{m,n}(\s).
\ee
Here, $i_{V}$ denotes interior derivation with respect to the vector $V$, and $\del_i$ is the $i$-th worldvolume tangent vector. When $Y\in\Omega^{p,n}$ we will also indicate an integrated version of this map using the same symbol:
\be
	\label{integrated bar map}
	\bar Y^{0,n}=\int d^p\s \bar Y^{0,n}(\s).
\ee
Even though we may omit the argument in (\ref{3:form correspondence}), it should be clear within context which of these maps is implied. We now show that this map generates the algebra modifications of the $p$-brane from its associated $D$ cocycle.

The Noether charges associated with a manifestly left invariant Lagrangian will be denoted $\bar Q_A$. One finds\footnote{``Bar" above $Q_A$ or $D_A$ is a definition, not an action of the map (\ref{3:form correspondence}). The notation indicates that $\bar Q_A$ and $\bar D_A$ naturally act upon elements in the image of this map.}:
\be
	\bar Q_{A}=\int d^{p}\s R_{A}{}^{M}P_{M}.
\ee
These charges are the phase space analog of the left generators (\ref{2:scalar left gen}). They satisfy the same algebra as the background superalgebra, but with the sign reversed:
\be
	[\bar Q_{A},\bar Q_{B}\}=-t_{AB}{}^{C}\bar Q_{C}.
\ee
This is the ``minimal algebra." In general, the $p$-brane Lagrangian is not manifestly left invariant (i.e. it is only symmetric up to a total derivative) due to quasi-invariance of the WZ term. Using the definitions of section \ref{sec:WZ terms as a double complex}, the variation of the WZ form is $Q_AB=-dW_A$. From this we have:
\bea
	Q_A\mathcal{L}&=&Q_A\mathcal{L}_{WZ}\\
	&=&\del_{i}w_{A}{}^{i}\nn,
\eea
where
\be
    \label{4:w related to W}
    w_{A}{}^{i}=-\frac{1}{p!}\tilde \e^{i_{p}...i_{1}i}W_{i_{1}...i_{p},A}
\ee
and $\tilde \e$ is the antisymmetric Levi-Civita symbol. Now, upon using the EL (Euler-Lagrange) equations:
\be
    \frac{\del \mathcal{L}}{\del Z^{M}}-\del_{i}\frac{\del \mathcal{L}}{\del(\del_{i}Z^{M})}=0,
\ee
we have identically:
\be
    Q_A\mathcal{L}=\del_{i}\bigg [Q_AZ^{M}\frac{\del\mathcal{L}}{\del(\del_{i}Z^{M})}\bigg ].
\ee
Hence, ``on-shell" there are conserved currents:
\bea
    \label{3:conserved current}
    \tilde q_{A}{}^{i}&=&Q_{A}Z^{M}\frac{\del \mathcal{L}}{\del(\del_{i}Z^{M})}-w_{A}{}^{i}\\
    \del_{i}\tilde q_{A}{}^{i}&=&0\nn.
\eea
The associated conserved charges are:
\be
	\label{6:cons charges}
	\widetilde {\bar Q}_{A}=\bar Q_{A}+\bar W_{A}.
\ee
Using (\ref{6:cons charges}), the $\widetilde {\bar Q}_{A}$ obey a modified version of the minimal algebra:
\be
    [\widetilde {\bar Q}_{A},\widetilde {\bar Q}_{B}\}=-t_{AB}{}^{C}\widetilde {\bar Q}_{C}+\bar M_{AB},
\ee
with
\be
    \label{6:M def}
    \bar M_{AB}=[\bar Q_{A},\bar W_{B}\}+[\bar W_{A},\bar Q_{B}\}+t_{AB}{}^{C}\bar W_{C}.
\ee
This is the algebra of conserved charges. Now define the special representative $M=sW$ of the $p$-brane $D$ cocycle. The definition of $\bar M$ given here agrees with that obtained from the bar map (\ref{integrated bar map}) acting upon $M$. We refer to both $M$ and $\bar M$ as ``anomalous terms." If we need to distinguish between the two, $\bar M$ will be referred to as the ``topological anomalous term", and $M$ as its ``superspace representation." The bar map ensures that elements in its image contain no time derivatives, or equivalently no dependence upon the phase space momenta. The anomalous term $\bar M$ then results from Poisson brackets involving at most one momentum variable, which leads to a simplified structure.

One verifies that:
\bea
	\bar M_{AB}&=&(-1)^p\int d^{p}\s M_{p\ldots 1,AB}(\s)\\
	&=&(-1)^p\int \Phi^* M_{AB}\nn,
\eea
where the map $\Phi$ embeds the spatial section of the worldvolume into superspace. We assume that the spatial section is a closed manifold. $\bar M_{AB}$ is therefore just a topological integral over the spatial section of the closed $p$-form $M_{AB}$. The result of the integral will be determined by the topology of the spatial section, and the class of the associated de Rham cohomology to which $M_{AB}$ belongs.

In prior literature the topological anomalous term was found to be proportional to the pullback of the $p$-form \cite{azc89}:
\be
    \G_{m_1\ldots m_p\a\b}dx^{m_1}\ldots dx^{m_p}.
\ee
A current associated with this form can be defined, and this current is conserved identically since the form is closed (see (\ref{4:conserved current}) below). However, this structure for the anomalous term assumes that integrals of the form:
\be
    \label{6:closed theta integral}
    \int d\s^{1}\del_{1}Y(\t)
\ee
vanish, where $Y$ is an arbitrary function. This amounts to the requirement that the fermionic directions (corresponding to the coordinates $\t$) must have trivial topology. The topological integrals of closed forms with $\t$ differentials and single valued coefficients must vanish in this case. However, recent work \cite{peeters03} suggests that for certain spaces more general than flat superspace, fermionic charges in the modified algebra are \textit{required} on the basis of Jacobi identities. In flat space it is consistent to set the fermionic charges to zero but in other spaces this can cause inconsistencies. Although we assume flat background spaces in this work, we will formally allow nonvanishing fermionic charges in order to see which features appear as a result. Since $\bar M$ is still derived from a closed form, the associated current is still conserved identically:
\bea
	\label{4:conserved current}
    m_{AB}^{i}&=&\tilde \e^{i_{p}...i_{1}i}M_{i_{1}...i_{p},AB}\frac{1}{p!}\\
    \del_{i}m_{AB}^{i}&=&\tilde \e^{i_{p}...i_{1}i}\del_i \del_{i_1}N_{i_{2}...i_{p},AB}\frac{1}{(p-1)!}\nn\\
    &=&0\nn.
\eea
There is no obvious reason to expect that it should be possible to incorporate the topological anomalous term $\bar M$ into the definition of an extended algebra. However, using its superspace representation $M$ we now show that this is indeed possible. In section \ref{sec:Application to the GS superstring} we will explicitly derive the extended algebras that result from the superstring anomalous term.
\begin{theorem}[extension]
The anomalous term of the Noether charge algebra defines an extension of the background superalgebra by an ideal. The resulting extended superalgebra is solvable.
\end{theorem}

First we need to show closure of the algebra. This requires that the anomalous term, and all brackets resulting from it, can be expressed using a finite number of new generators. To find the extended algebra one could investigate the Poisson brackets of $\widetilde{\bar Q}_A$ and $\bar M_{AB}$. However, one can equivalently use the double complex. In this case the anomalous term is represented by a set of superspace forms $M_{AB}$ (one for each bracket of the minimal algebra). The minimal algebra is generated by the left generators $Q_A$ of the double complex. Define the modified left generators as:
\be
	\widetilde Q_A=Q_A+W_A.
\ee
The Poisson bracket algebra generated by $\widetilde {\bar Q}_A$ and $\bar M$ is found to be the same as the ``operator-form" algebra generated by $\widetilde Q_A$ and $M$ (forms are assumed to commute with other forms). We therefore use the operator-form algebra since it is more convenient. If required, the Poisson bracket algebra can be obtained by replacing all generators with barred ones in the operator-form algebra. Let $\mathcal{G}=\{Q_A\}$ denote the minimal algebra, and $\mathcal{\widetilde G}=\{\widetilde Q_A,\Sigma_A\}$ denote the full algebra that is assumed to result by addition of the anomalous term. Now consider the following schematic representation of the action of the left generators on forms\footnote{For this proof we will assume for brevity that $\mathcal{G}$ is the standard superalgebra. The same principles are also valid in the case where $\mathcal{G}$ is one of the extended superalgebras (e.g. those of section \ref{sec:Extended algebras}).}:
\begin{center}
$\begin{array}{llllll}
    x &\rightarrow & \t &\rightarrow & $const $ \rightarrow & 0\\
    dx &\rightarrow & d\t &\rightarrow & 0. &
\end{array}$
\end{center}
If a form has coefficients with a polynomial structure then each action of $Q_A$ brings it closer to annihilation. The requirements of Lorentz invariance and fixed dimensionality (see section \ref{sec:Anomalous term cohomology}) ensure that all valid forms have this polynomial structure. It follows that the anomalous term will be annihilated by the left generators in a finite number of steps. There is then a stepwise process to define the extended algebra. At the first step we may factor out any Lorentz invariant tensors from $M_{AB}$ (which become new structure constants). The remaining form is then written in terms of a minimal set of independent closed forms $\Sigma_A$, which become new generators of the algebra. The $\Sigma_A$ commute with themselves and satisfy:
\be
	\label{modifed operator with form bracket}
	[\widetilde Q_A,\Sigma_B\}=[Q_A,\Sigma_B\}
\ee
since $W_A$ commutes with $\Sigma_B$. We then act again with the $Q_A$ and introduce new generators to deal with any forms that cannot be written in terms of those generators previously defined. By the above annihilation argument it follows that this process is finite. That is, there will be a finite number of new generators. The resulting algebra has the structure:
\bea
    \ [\widetilde Q,\widetilde Q]&\subset& \widetilde Q\oplus\Sigma\\
    \ [\widetilde Q,\Sigma]&\subset& \Sigma\nn
\eea
The second line shows that $\Sigma$ is an ideal of the new algebra. The algebra $\mathcal{\widetilde G}$ is said to be \textit{solvable} if:
\be
    (\mathsf{Ad}_\mathcal{\widetilde G})^m(\mathcal{\widetilde G})=0
\ee
for some finite integer $m$, where $\mathsf{Ad}_\mathcal{\widetilde G}$ is the adjoint action. The minimal algebra $\mathcal{G}$ is solvable. The annihilation argument shows that $\mathcal{\widetilde G}$ is also solvable, since the action of $\mathcal{\widetilde G}$ annihilates the new generators in a finite number of steps.

This shows that the new algebra closes, however to show that $\mathcal{\widetilde G}$ is a valid \textit{superalgebra} we must also show that the super-Jacobi identities are satisfied. There are four cases to test. The first is:
\be
    (-1)^{AC}[\widetilde Q_{A},[\widetilde Q_{B},\widetilde Q_{C}\}\}+\mathsf{cycles},
\ee
where ``$\mathsf{cycles}$" indicates the terms obtained from two repetitions of the cycling $A\rightarrow B\rightarrow C$. Using $M=sW$ one can show that this reduces to:
\be
    (-1)^{AC}t_{BC}{}^Dt_{AD}{}^E Q_{E}+\mathsf{cycles},
\ee
which vanishes since the original structure constants satisfy the Jacobi identity. The second case is:
\be
    (-1)^{AC}[\widetilde Q_{A},[\widetilde Q_{B},\Sigma_{C}\}\}+\mathsf{cycles}.
\ee
By (\ref{modifed operator with form bracket}) it is valid to replace $\widetilde Q$ by $Q$ in the above expression since $\Sigma$ is an ideal. The Jacobi identity is then identically satisfied since it reflects an action of the minimal algebra. The final two cases:
\be
    (-1)^{AC}[\widetilde Q_{A},[\Sigma_{B},\Sigma_{C}\}\}+\mathsf{cycles}\\
    (-1)^{AC}[\Sigma_{A},[\Sigma_{B},\Sigma_{C}\}\}+\mathsf{cycles}\nn
\ee
are trivially satisfied. The Jacobi identity therefore holds, and $\mathcal{\widetilde G}$ is an extended superalgebra.

\subsection{The algebras of right generators and constraints}
\subsubsection{The algebra of right generators}

The right generators and their algebra are modified in a similar way to the left generators. The minimal right generators for the phase space are:
\be
	\label{3:unmodified right gen def}
	\bar D_{A}=L_{A}{}^{M}P_{M}.
\ee
The $\bar D_{A}$ satisfy the minimal algebra:
\be
	[\bar D_{A}(\s),\bar D_{B}(\s')\}=\d(\sa-\sa')t_{AB}{}^{C}\bar D_{C}(\s).
\ee
If a WZ term is added to the NG action, the new momenta are related to the NG action momenta $P^{(NG)}_{M}$ via:
\be
    \label{3:momenta extra term}
    P_{M}=P^{(NG)}_{M}+\bar B_{M}.
\ee
The WZ term of the Lagrangian may be written in terms of $\bar B$ as:
\be
    \label{3:Adef}
    \mathcal{L}_{WZ}=\dot Z^{M}\bar B_{M}.
\ee
We define the modified right generators for the phase space such that they are constructed from the NG momenta:
\bea
    \label{3:modified right gen}
    \widetilde {\bar D}_{A}&=&L_{A}{}^{M}P^{(NG)}_{M}\\
    &=&\bar D_{A}-\bar B_{A}\nn.
\eea
This is motivated by the modification of the standard superspace action constraints in the presence of the WZ term \cite{azc91}, and the relation of constraints to right generators (see section \ref{3:sec:constraints}). Again, the components of $\bar B$ contain no time derivatives. Thus, the modification to the right generators for the phase space contains no momentum dependence (just as in the left generator case). If one imposes a condition that $B$ must be single valued then the modified algebra derives from $H$:
\be
    \label{3:right gen modified algebra}
    [\widetilde {\bar D}_{A}(\s),\widetilde {\bar D}_{B}(\s')\}=\d(\sa-\sa')[t_{AB}{}^{C}\widetilde {\bar D}_{C}-\bar H_{AB}](\s).
\ee
This shows that the result stated in \cite{azc91} for the standard superspace action also holds for extended superspace actions. The bar map is again seen to commute with the bracket operation:
\bea
    \label{6:Fdef}
    \d(\sa-\sa')\bar H_{AB}(\s)&=&[\bar D_{A}(\s),\bar B_{B}(\s')\}+[\bar B_{A}(\s),\bar D_{B}(\s')\}\\
    &&-\d(\sa-\sa')t_{AB}{}^{C}\bar B_{C}(\s)\nn.
\eea

\subsubsection{The algebra of constraints}
\label{3:sec:constraints}

The $p$-brane action (\ref{3:p-brane action}) yields constraint equations for the phase space variables. That is, equations of the form:
\be
	\label{3:constraint defined}
	C_{M}(Z,P)=0
\ee
for some functions $C_{M}$, which reduce to identities once the definitions (\ref{3:momentadef}) of momenta are used.
This results in a reduction of phase space. For the content of this paper it will only be necessary to find (not eliminate) the constraints.

Evaluating $\frac{\del L}{\del \dot x^{m}}$ and $\frac{\del L}{\del \dot \t^{\m}}$ for the NG action one finds:
\bea
    P^{(NG)}_{m}&=&-\gmh g^{0i}L_{i}{}^{a}\eta_{am}\\
    P^{(NG)}_{\m}&=&-\half(\G^{n}\t)_{\m}P^{(NG)}_{n}\nn.
\eea
One thus identifies the fermionic constraints of the NG action as:
\bea
    C_{\a}&=&\d_{\a}{}^{\m}P^{(NG)}_{\m}+\half(\G^{n}\t)_{\m}P^{(NG)}_{n}\\
    &=&L_{\a}{}^{M}P^{(NG)}_{M}.\nn
\eea
Comparing with (\ref{3:unmodified right gen def}) we see that these are just the odd, minimal right generators for the phase space. The $C_{\a}$ thus satisfy the algebra:
\be
    \{C_{\a}(\s),C_{\b}(\s')\}=\d(\sa-\sa')t_{\a\b}{}^{A}\bar D_{A}(\s).
\ee
Upon the addition of a WZ term, the momenta (including those associated to new coordinates) pick up the extra terms $\bar B_M$ as in (\ref{3:momenta extra term}). It will be assumed that the background superspace is either standard superspace, or an extension of standard superspace by an ideal (e.g. the superalgebras of section \ref{sec:Extended algebras}). We find that the constraints $\widetilde C_{A}$ in the presence of the WZ term can then be written:
\be
    \label{6:fermionic mod}
    \widetilde C_{A}=L_{A}{}^{M}(P_{M}-\bar B_{M}),\qquad A\neq a.
\ee
Details of the calculation may be found in appendix \ref{sec:app:algebra conditions}. Thus, the constraints $\widetilde C_{A}$ (where $A\neq a$) are the modified right generators for the phase space (\ref{3:modified right gen}), and their algebra is the same:
\be
	[\widetilde C_{A}(\s),\widetilde C_{B}(\s')\}=\d(\sa-\sa')[t_{AB}{}^{C}\widetilde {\bar D}_{C}-\bar H_{AB}](\s).
\ee
Note that although there is no constraint $\widetilde C_{a}$, $\widetilde {\bar D}_{a}$ can still appear on the RHS.

The constraint surface must be invariant under the action of the Noether symmetries of the action. The constraints must therefore be left invariant in the sense:
\be
    [\bar {Q}_{A},C_{\b}(\s)\}\approx 0,
\ee
where $\approx$ means ``equal on the constraint surface." For the NG action this is an example (in PB form) of the commutativity of the left and right actions. When the WZ term is added, this condition must continue to hold (i.e. upon replacing $\bar {Q}_{A}$ and $C_{A}$ by their modified counterparts). In fact, if one assumes that $W$ is single valued then one can use the descent equation $sB=-dW$ to show:
\be
    [\widetilde {\bar Q}_{A},\widetilde {\bar D}_{B}(\s)\}=0.
\ee
This generalizes a result in \cite{azc91} for the standard background to the case of extended backgrounds. Since the constraints are a subset of the modified right generators, their left invariance is guaranteed by the double complex cohomology. Furthermore, since the equation $sB=-dW$ is preserved by the gauge transformations, the left invariance of the constraints is independent of the gauge.

\subsection{Cohomology of algebra modifications}
\label{sec:Anomalous term cohomology}

We are now in a position to determine how gauge freedom affects the algebras of left/right generators. Before proceeding however, we need to establish some facts about the $D$ cohomology of $H$. First let us review why the equation (\ref{3:H def}) defining $H$ takes the form it does. $H$ must have the following properties:
\subsubsection{Properties of $H$}
\begin{itemize}
\label{Properties of H}
\item
$H$ is closed.
\item
$H$ is left invariant.
\item
dim $H=p+1$.
\item
$H$ is Lorentz invariant.
\end{itemize}
In standard superspace, $H$ is the \textit{unique} $p+2$ form (up to a constant of proportionality) with the properties \ref{Properties of H} \cite{azc89-2}. Furthermore, it is a nontrivial CE cocycle. In the double complex construction this implies that in standard superspace, $H$ is the unique Lorentz invariant element of $H^{p+2,0}$ with dimension $p+1$. One can verify that the last two items in the list \ref{Properties of H} are preserved by the operators $d$ and $s$. We conclude that Lorentz invariance and dimensionality $p+1$ must be a property of \textit{all} elements of the double complex (including potentials and gauge transformations). The exactness of $s$ means that the $D$ cohomology of the single complex is equal to the de Rham cohomology of the first column of the double complex. Since we should restrict ourselves to Lorentz invariant forms of dimension $p+1$, by the uniqueness of $H$, this cohomology is equal to the field of scalars we are using (the constant of proportionality multiplying $H$ labels the class).

The uniqueness of $H$ implies that the modification to the right generator algebra $\bar H$ is also unique. It is gauge invariant, and even independent of the background superalgebra used (since the same definition of $H$ is always used). Note however that the right generator algebra obtained in an extended background can be different to the right generator algebra obtained in standard superspace (even though the modification is the same) because then the \textit{minimal} algebra that we start with is already different.

The left generator algebra is less straightforward. Note that due to nilpotency of the operators, moving twice in any one direction on the tic-tac-toe box gives zero. An interesting consequence of this is that the gauge freedom in the WZ term (resulting from $\psi$, $C'$) has no effect upon the anomalous term $M$. Note however that using a different background superspace will not only change the minimal algebra but can also change the modification $M$ (since the descent equations may have different solutions). We note that left invariant WZ terms can only be constructed in such extended superspace backgrounds.

The result of main interest is that the topological anomalous term $\bar M$ is not gauge invariant. Using (\ref{3:delta W}):
\bea
    \label{Delta M=-sd lamda}
    \D M&=&s\D W\\
    &=&sd\lambda\nn.
\eea
Although $\D M$ is a $D$ coboundary, it need not be exact. $\D \bar M$ can therefore be nonzero in the presence of nontrivial topology (just as $\bar M$ can be). How much freedom do we have? At first it seems that we have full gauge freedom at our disposal, but in practice the requirements of Lorentz invariance and correct dimensionality are restrictive. In section \ref{sec:Application to the GS superstring} we will see that in the case of the string, these requirements on the gauge fields reduce the freedom in the anomalous term down to a single, global degree of freedom. A corresponding free constant parameterizes the ``spectrum of algebras" obtained from the process.

Identifying gauge freedom in the anomalous term forces us to reevaluate its mathematical nature. Since there is an orbit of gauge equivalent representatives, and there is no natural basis upon which to fix a gauge, one can no longer speak of ``the" anomalous term if one defines it as a particular form or modified left generator algebra. In order that the anomalous term be a well defined object it must be defined as an entire $D$ cohomology class $[M]$. We have already seen that the representatives $M$ of this class are $D$ cohomologous to $H$. Since $s$ is exact, this correspondence is a \textit{bijection} between the $H^{p+2,0}$ and $H^{p,2}$ cohomologies to which $H$ and $M$ belong. That is, to each cohomology class $[H]\in H^{p+2,0}$ is associated a unique class $[M]\in H^{p,2}$ of the \textit{same triviality}, and vice versa. The nature of the resulting class $[M]$ depends on the background space being used.

First consider standard superspace. Since the class $[H]$ is unique and nontrivial, $[M]$ must also be unique and nontrivial. The classes $[M]$ must be labeled by a single proportionality constant belonging to the field of scalars (just as the classes $[H]$ are). The difference between $[H]$ and $[M]$ is that $[H]$ consists of $H$ only; there are no \textit{coboundaries} for the $H^{p+2,0}$ cohomology. In general there \textit{are} coboundaries for the $H^{p,2}$ cohomology; they are precisely the $\lambda$ gauge transformations (and we will see that explicit, nonvanishing examples of such gauge transformations do exist). The $D$ cocycle of the $p$-brane therefore has a set of equivalent representatives belonging to $\Omega^{p,2}$; it is this full set which makes the anomalous term a well defined object.

If an extended superspace is used then $H$ is a $D$ coboundary. Based on the historical derivation, one might argue that in this case the anomalous term should not even exist (since a manifestly left invariant WZ term is possible). However, from the cohomology point of view the anomalous term should consist of all possible modifications to the Noether charges that are consistent with charge conservation. In the double complex construction, charge conservation is guaranteed by the descent equations. The anomalous term $[M]$ therefore becomes the space of $D$ coboundaries within $\Omega^{p,2}$. This is identically equal to the representatives $\D M$ resulting from the $\lambda$ gauge transformations. Note that $D$ coboundaries need not be exact; it is therefore possible to obtain nonzero topological integrals for $\bar M$ even in the case of a manifestly left invariant WZ term.

We summarize with the following:
\begin{theorem}[cohomology]
The anomalous term is the restriction of $H^{p,2}$ to forms that are $D$ cohomologous to $H$.
\end{theorem}
\begin{theorem}[uniqueness]
In the standard background, the anomalous term is the unique, Lorentz invariant, $D$ nontrivial class of dimensionality $p+1$ (uniqueness is up to a proportionality constant).
\end{theorem}

From the second of these we conclude that in standard superspace it is possible to find the anomalous term without solving descent equations. If a single $D$ nontrivial representative within $H^{p,2}$ can be found then the entire anomalous term will be generated by the $\lambda$ gauge transformations. This class is unique (up to the constant of proportionality which labels the classes). In superspaces which allow manifestly left invariant Lagrangians, the anomalous term is the set of $D$ coboundaries generated by the $\lambda$ gauge transformations.

Note that the above arguments apply only to the superspace representation $M$ of the anomalous term. The associated topological anomalous term $\bar M$ may vanish for topological reasons separate from $D$ cohomology. For example, if we choose to compactify \textit{no} dimensions, or if the brane does not ``wrap", then topological integrals such as $\bar M$ must identically vanish. If we compactify only \textit{some} dimensions then we may find that in standard superspace there do exist gauges in which the topological anomalous term vanishes, since a gauge transformation may shift the form $M$ into a trivial sector of the cohomology of the spatial section. We will see an explicit example of this in section \ref{sec:Application to the GS superstring}.

To summarize, the $p$-brane has an associated $D$ cocycle defined by the representative $H\in H^{p+2,0}$. The Noether charge algebra can be modified by a topological anomalous term deriving from cocycle representatives $M\in H^{p,2}$. The representatives are not unique due to the presence of $\lambda$ gauge transformations of the cocycle. These transformations themselves represent topological integrals which can be nonzero. The anomalous term is well defined as a cohomology class, where elements related by $\lambda$ gauge transformations are to be considered equivalent. Since each representative of the anomalous term defines an extended supertranslation algebra, each algebra in the spectrum can be considered as being equivalent from a $D$ cohomology point of view.

It is interesting to note that all the cocycle representatives of ghost degree two or less have physical interpretations:
\begin{itemize}
\item
H measures the modification to the right generator algebra.
\item
sB measures the left variation of the WZ term.
\item
sW measures the modification to the left generator algebra.
\end{itemize}
One may ask if any other representatives are significant. The only one remaining in the case of the string is the ghost degree three element $sN$. Consider the following modified algebra\footnote{We present this for the sake of interest only since we have no physical interpretation for modified algebras resulting from $N$.}:
\be
	[Q_{A},Q_{B}\}=-t_{AB}{}^{C}Q_{C}+N_{AB}.
\ee
One finds that the Jacobi identity of this algebra is generated by $sN$:
\be
	(-1)^{AC}sN_{ABC}=(-1)^{AC}[Q_A,[Q_B,Q_C\}\}+\mathsf{cycles}.
\ee
$sN$ therefore determines whether or not $N$ can define an extended superalgebra. Based on the cocycle triviality arguments we conclude that $N$ defines extensions of extended backgrounds, but not of the standard background. Applying the same argument to $M$ (and using $sM=0$), we verify the claim of section \ref{sec:The algebra of left generators} that $M$ generates extensions of both standard and extended backgrounds.

We finally note that the argument which shows that $H^{p+2,0}$ is unique in standard superspace implies the same for $H^{0,p+2}$. That is, the class containing $sB^{0,p+1}$ consists of one element. The components of $sB^{0,p+1}$ must also be proportional to those of $H$ since the construction of a nontrivial representative in $H^{0,p+2}$ has the same mathematical content as the construction of a nontrivial representative in $H^{p+2,0}$.

\section{Application to the GS superstring}
\label{sec:Application to the GS superstring}

To illustrate the above formalism we consider the case of the GS superstring. After presenting the action, the modified algebras of the left/right generators are found. The effect of the cocycle gauge transformations is then investigated.

\subsection{Superstring actions}

We wish to study the effects that the following may have upon the results:
\begin{itemize}
\item
Extending the background superspace (in order to allow manifestly symmetric WZ terms to be used).
\item
Changing the WZ term.
\end{itemize}
For this purpose we use an action that has free parameters (``switches"). The action can be used in the standard superspace background and also on the two extended ones of section \ref{sec:Extended algebras}. The switches allow one of three WZ terms to be used, or alternatively no WZ term at all. The action is:
\bea
	\label{3:action}
	S_{k,s,\bar s}&=&-\int d^{2}\s \sqrt{-g}\ \bigg [1-\frac{k}{2}\e^{ij}\bigg (\tb\G_{i}\del_{j}\t-s\bigg [1-\frac{\bar s}{2}\bigg ]\del_{i}\t^{\m}\del_{j}\p_{\m}\\
		&&-s\bar s\del_{i}y_{n}\del_{j}x^{n}\bigg )\bigg ].\nn
\eea
The switches $k$, $s$ and $\bar s$ are restricted to the following values:\\
\\
$\begin{array}{ll}
k=\{-1,0,1\} & $controls the existence and sign of the WZ term.$\\
s=\{0,1\} & $switches on a manifestly invariant WZ term.$\\
\bar s=\{0,1\} & $controls the type of invariant WZ term.$
\end{array}$\\
\\

$k=0$ gives the NG action. For $k\neq 0$ we have three possibilities.
\begin{itemize}
\item
$s=0$ gives the standard WZ term on standard superspace. This results in the standard $\k$ symmetric GS superstring action. The corresponding Lagrangian is only left invariant up to a total derivative.
\item
$(s,\bar s)=(1,0)$ gives a manifestly left invariant WZ term that exists on the superspace of the Green algebra \cite{siegel94,bergshoeff95}. The resulting action can be brought to the form:
\be
	S_{k,1,0}=-\int d^{2}\s \sqrt{-g} \bigg [1+\frac{k}{2}\e^{ij}(L_{i}{}^{\a}L_{j\a})\bigg ],
\ee
showing clearly the manifest left invariance. The WZ 2-form in this case is:
\be
    B=\frac{k}{2}L^{\a}L_{\a}.
\ee
\item
$(s,\bar s)=(1,1)$ gives another manifestly left invariant WZ term that exists on the superspace of the extended Green algebra \cite{chrys99}. In this case:
\be
	\label{2 term LI WZ form}
    B=-\frac{k}{2}L^{a}L_{a}+\frac{k}{4}L^{\a}L_{\a}.
\ee
\end{itemize}

\subsection{Constraint and right generator algebras}
The action (\ref{3:action}) yields the bosonic momentum:
\be
    P_{m}=-\gph g^{0i}L_{i}{}^{a}\eta_{am}-\frac{k}{2}\tgdotl{m}.
\ee
The momenta other than $P_{m}$ can be written in terms of $P_{m}$ and $Z^{M}$. These equations are then written in the form of constraints on phase space\footnote{These are the ``modified" constraints of the general section. We have dropped the tilde since we are no longer considering the NG and GS actions separately.}:
\bea
    C_{\m}&=&P_{\m}
        +\half \gtu{m}{\m}P_{m}
        +\frac{k}{2}L_{1}^{a}\gtl{a}{\m}
        +\frac{k}{4}s\bar s(\G^{n}\t)_{\m}\del_{1}y_{n}\\
        &&-\frac{sk}{2}\bigg [1-\frac{\bar s}{2}\bigg ]\del_{1}\p_{\m}\nn\\
    C^{m}&=&P^{m}-\frac{k}{2}s\bar s\del_{1}x^{m}\nn\\
    C^{\m}&=&P^{\m}-\frac{sk}{2}\bigg [1-\frac{\bar s}{2}\bigg ]\del_{1}\t^{\m}\nn.
\eea
The 1-form $\bar B$ is found to be:
\bea
    \bar B_{m}&=&-\frac{k}{2}\tgdotl{m}-\frac{k}{2}s\bar s\del_{1}y_{m}\\
    \bar B_{\m}&=&\frac{ks}{2}\bigg [1-\frac{\bar s}{2}\bigg ]\del_{1}\p_{\m}-\frac{k}{2}\del_{1}x^{m}\gtl{m}{\m}\nn\\
    \bar B^{m}&=&\frac{k}{2}s\bar s\del_{1}x^{m}\nn\\
    \bar B^{\m}&=&\frac{ks}{2}\bigg [1-\frac{\bar s}{2}\bigg ]\del_{1}\t^{\m}\nn.
\eea
In standard superspace, $C_{\m}$ coincide with the right generators for the phase space, but for the extended algebras it is the linear combinations of section \ref{3:sec:constraints} that generate the right action. These are:
\be
    \label{3:rightwithA}
    C_{A}=L_{A}{}^{M}P_{M}-\bar B_{A},
\ee
where $\bar B_{A}=L_{A}{}^{M}\bar B_{M}$.

For the string we require D=(3, 4, 6, 10) \cite{evans88}. The Fierz identity becomes:
\be
    \label{2:simplifying}
    \G^{a}{}_{(\a\b}\G_{a\d)\e}=0.
\ee
Using this, the Poisson bracket algebra of the constraints is found to be:
\bea
    \label{3:rightalg}
    \{C_{\a}(\s),C_{\b}(\s')\}&=&\d(\sa-\sa')(\G^{a}{}_{\a\b}\widetilde{\bar D}_{a}+k\G_{a\a\b}L_1{}^a)(\s)\\
    \ [C^{a}(\s),C_{\b}(\s')]&=&-\d(\sa-\sa')\G^a{}_{\b\g}C^{\g}(\s)\nn,
\eea
with all other brackets vanishing. The second bracket is an example illustrating the fact that although the modification $H^a{}_\b$ vanishes, the associated constraint bracket is nonzero for the extended Green algebra because the minimal algebra has a noncentral generator $\Sigma^a$. Note that the constraint $C^a$ does not exist on standard or Green superspaces, and in these cases only the first bracket is present.

The algebra of right generators is slightly more general than (\ref{3:rightalg}) since there is a generator $D_{a}$ that is not reflected as a constraint. Using the bar map (\ref{3:form correspondence}) and the components of $H$:
\be
    H_{c\b\a}=k\G_{c\b\a}
\ee
we obtain:
\bea
    \bar H_{\a\b}&=&-k\G_{a\a\b}L_{1}{}^{a}\\
    \bar H_{a\b}&=&k(\G_{a}\del_{1}\t)_{\b}\nn
\eea
as the only nonzero components of the modification. The first of these is seen to agree with the first bracket of (\ref{3:rightalg}). The second is not present in the constraint case.

\subsection{Left generator algebra}

\subsubsection{Standard superspace action}
\label{sec:standard superspace action}

Let us find a representative of the anomalous term by solving the descent equations. First, using the Fierz identity one finds for the variation of the WZ form:
\bea
    \label{6:std variation}
    Q_{\a}B&=&-\frac{k}{2}L^{b}(\G_{b}d\t)_{\a}\\
    &=&-\frac{k}{2}d\bigg [(dx^{b}-\sixth d\bar\t\G^b\t)(\G_{b}\t)_{\a}\bigg ]\nn.
\eea
The bosonic symmetries are manifest (i.e. $Q_{a}B=0$). Thus:
\be
    \label{6:std solution for W}
    W=\frac{k}{2}e^{\a}(dx^{b}-\sixth d\bar\t\G^b\t)(\G_{b}\t)_{\a}
\ee
is a solution for the potential $W$. Evaluating $M=sW$ and using the Fierz identity we find that all $\t$ dependence is lost:
\be
\label{6:standard modification}
    M_{\a\b}=kdx^{m}\G_{m\a\b},
\ee
with all other components vanishing. Using the map (\ref{integrated bar map}) we then find $\bar M$:
\be
	\label{6:standard physical modification}
	\bar M_{\a\b}=-k\int d\s^{1}\del_{1}x^{m}\G_{m\a\b}.
\ee
This integral can be nonzero whenever the spatial section has nontrivial topology in the bosonic sector. It is equivalent to the previously known result \cite{azc89} except that we have not needed to assume trivial fermionic topology. One of the new points is that (\ref{6:std variation}) determines $W$ only up to a gauge transformation (which we have called $\lambda$). The resulting anomalous term $M$ is not gauge invariant under such transformations. In fact, we now show that if fermionic topology is trivial then the topological anomalous term (\ref{6:standard physical modification}) is gauge equivalent to zero.

The following gauge field satisfies the conditions of Lorentz invariance and dimensionality $p+1=2$:
\be
    \label{6:explicit lambda}
    \lambda=-ke^a x^{b}\eta_{ab}.
\ee
Let us find its effect upon the solutions (\ref{6:std solution for W}) and (\ref{6:standard modification}) for $W$ and $M$. Firstly:
\bea
	\D W&=&d\lambda\\
	&=&-ke^a dx^{b}\eta_{ab}\nn.
\eea
Using $\D M=sd\lambda$ we then find:
\bea
    \D M_{\a\b}&=&-kdx^{m}\G_{m\a\b}\\
    \D M_{a \b}&=&-\frac{k}{2}(\G_{a}d\t)_{\b}\nn\\
    \D M_{ab}&=&0\nn.
\eea
We see that $\D M_{\a\b}$ is closed but not exact whenever $dx^{m}$ is. Now, the de Rham nontriviality of $dx^{m}$ is the condition for which the original representative (\ref{6:standard physical modification}) is nonzero. Therefore, in this case the gauge transformation $\D\bar M$ is nonzero whenever $\bar M$ itself is.

After the gauge transformation, the alternative representative $M'$ is:
\be
    \label{6:mirror symmetry gauge}
    M'_{a \b}=-\frac{k}{2}(\G_{a}d\t)_{\b}.
\ee
We have thus traded nonzero $M_{\a\b}$ for nonzero $M_{a\b}$. However, when converted to the topological anomalous term this becomes a topological $\t$ integral of the type (\ref{6:closed theta integral}):
\be
    \label{6:QP top charge}
    \bar M'_{a \b}=\frac{k}{2}\int d\s^{1}(\G_{a}\del_{1}\t)_{\b}.
\ee
Therefore, even if the standard quasi-invariant Lagrangian is used, when fermionic topology is trivial, the topological charge algebra is gauge equivalent to the minimal algebra.

The more interesting case occurs when nontrivial fermionic topology is formally allowed. In this case, the integral (\ref{6:QP top charge}) can be nonzero. Let us repeat the above procedure using instead the associated one parameter family of gauge transformations parameterized by a constant $a$:
\be
    \label{6:gauge xfm one parameter}
    \lambda=-ake^a x^{b}\eta_{ab}.
\ee
First we show that this is in fact the most general gauge transformation. There are two more possibilities for $\lambda$ with the correct Lorentz and dimensionality properties. The first is:
\be
    \label{6:redundant gauge xfm}
    \lambda'=-\frac{ak}{2}x^{a}\bar e\G_{a}\t.
\ee
Defining $\D'W=d\lambda '$, one can verify that although $\D W$ differs from $\D' W$, the algebra modifications $\D M$ and $\D'M$ are the same. In the context of this paper it is the algebra itself that is important, not any particular representation of its generators. The transformation (\ref{6:redundant gauge xfm}) is therefore equivalent to (\ref{6:gauge xfm one parameter}). The only other possibilities appear to be gauge fields of the form:
\be
	\lambda''=\bar e\G^{a_1. . .a_b}\t \bar\t\G_{a_1. . .a_b}\t,
\ee
where $b$ is such that $\G_{a_1. . .a_b\a\b}$ is antisymmetric. These transformations leave $M$ invariant, and are hence redundant. We may therefore take (\ref{6:gauge xfm one parameter}) as the most general transformation. Applying this to the representative (\ref{6:standard physical modification}) one finds the equivalence class $[\bar M]$ of topological anomalous terms, with representatives parameterized by the gauge parameter $a$:
\bea
	\ [\bar M]_{\a\b}&=&-(1-a)k\G_{m\a\b}\int d\s^{1}\del_{1}x^{m}\\
	\ [\bar M]_{a \b}&=&\frac{ak}{2}\int d\s^{1}(\G_{a}\del_{1}\t)_{\b}\nn.
\eea

By introducing appropriately defined new generators we now show that this anomalous term generates extended superalgebras. The new generators are simply the topological charges:
\bea
    \bar \Sigma^{a}&=&\frac{k}{2}\int d\s^{1}\del_{1}x^{a}\\
    \bar \Sigma^{\g}&=&\frac{k}{2}\int d\s^{1}\del_{1}\t^{\g}\nn.
\eea
Note that $\bar \Sigma^{a}$ and $\bar \Sigma^{\g}$ are nonzero only when the associated superspace dimension is compact and the spatial section of the string wraps around it. Upon adding these to the set of conserved charges:
\bea
    \widetilde {\bar Q}_{\a}&=&R_{\a}{}^{M}P_{M}-\frac{k}{2}\int       d\s^{1}(\del_{1}x^{m}-\sixth\del_{1}\bar\t\G^{m}\t)\gtl{m}{\a}\\
    \widetilde {\bar P}_{a}&=&R_{a}{}^{M}P_{M}+ak\int d\s^{1}\del_{1}x^{m}\eta_{ma}\nn,
\eea
we then obtain the following algebra under Poisson bracket:
\bea
    \label{6:gauge fixed algebra}
    \{\widetilde {\bar Q}_{\alpha},\widetilde {\bar Q}_{\beta}\}&=&-\Gamma^{b}{}_{\alpha\beta}\widetilde {\bar P}_{b}
        -2(1-a)\Gamma_{b}{}_{\alpha\beta}\bar \Sigma^{b}\\
    \ [\widetilde {\bar Q}_{\a},\widetilde {\bar P}_{b}]&=&-a\Gamma_{b\a\g}\bar \Sigma^{\g}\nn\\
    \ [\widetilde {\bar Q}_{\a},\bar \Sigma^{b}]&=&-\half \Gamma^b{}_{\a\g}\bar \Sigma^{\g}\nn.
\eea
We will check that the Jacobi identity is satisfied. The only nontrivial possibility is:
\be
	[\widetilde {\bar Q}_{\a},\{\widetilde {\bar Q}_{\b},\widetilde {\bar Q}_{\g}\}]+\mathsf{cycles}=3\G^{a}{}_{(\a\b}\G_{a\g)\d}\bar\Sigma^{\d},
\ee
which vanishes by the Fierz identity. We note three special cases:
\begin{itemize}
\item
For $a=1$ the extra generator $\bar \Sigma^{a}$ is redundant and may be excluded since it appears nowhere on the RHS of a bracket. We then recover the Green algebra\footnote{Negative signs relative to the background superalgebras are expected due to the use of operators instead of superalgebra generators. Redefinition of the operators with a sign reversal gives the background superalgebra.}.
\item
For $a=\half$ we rescale $\bar \Sigma^\a$ with a factor of $\half$ and recover the extended Green algebra.
\item
Turning off the gauge transformation altogether results in a variant in which $\widetilde {\bar P}_{a}$ is central. The structure of this algebra is of the type considered in \cite{peeters03}:
\bea
    \{Q,Q\}&\sim &P+P'\\
    \ [Q,P']&\sim &\Sigma\nn.
\eea
\end{itemize}
An important point is that the spectrum (\ref{6:gauge fixed algebra}) cannot be obtained by simply rescaling the known algebras. It is therefore a generalization which yields new superalgebras.

We see that the outcome of the construction is a spectrum of superalgebras parameterized by a free constant. The algebras are constructed by identifying topological charges with new superalgebra generators. One can then decompose the ideal arising from the topological anomalous term. The anomalous term, which is the modification to the Noether charge algebra in the presence of the nontrivial WZ term, contains a gauge freedom. The free constant of the algebra represents the part of the gauge freedom which is consistent with Lorentz invariance and dimensionality requirements. The spectrum of algebras contains the three superalgebra extensions that have so far been associated with the string. We emphasize two departures from prior literature that were required:
\begin{itemize}
\item
Since representatives of the anomalous term are not gauge invariant, it is well defined only as an entire cohomology class. A free constant parameterizes the class.
\item
The fermionic extensions of the superalgebra resulting from the anomalous term are topological integrals. If we formally allow nontrivial fermionic topology, these charges can be \textit{physically} realized. However, regardless of topological considerations, the extended superalgebras generated by the mechanism can always be \textit{abstractly} realized in the operator-form representation.
\end{itemize}

\subsubsection{Extended superspace actions}
\label{4:subsubsec:manifest examples}

The motivation behind using an extended background superspace was to enable a manifestly left invariant WZ term to be used. The left invariant WZ form is generated by a $\psi$ gauge transformation on the standard WZ form:\\
\\
$\bar s=0$:
\bea
    \D B&=&-d\psi\\
    &=&\frac{k}{2}d\t^{\m}d\p_{\m}\nn.
\eea
$\bar s=1$:
\bea
    \D B&=&-d\psi\\
    &=&-\frac{k}{2}dx^{m}dy_m+\frac{k}{4}d\t^{\m}d\p_{\m}\nn.
\eea
The manifest left invariance of the $s=1$ action allows us to choose vanishing components for $W$. $M=0$ is then a representative of the anomalous term. As expected, $M$ is therefore $D$ trivial for the extended superspace actions as a result of manifest left invariance.

When using extended superspaces it is just as valid to use the standard WZ term as any other one (since they are gauge equivalent). Let us therefore consider using the standard action on an extended superspace. In this case the extra available coordinates still trivialize the anomalous term. For example, in the case of the Green algebra one can modify $W$ from (\ref{6:std solution for W}) using:
\bea
    \D W&=&d\lambda\\
    &=&-\frac{k}{2}e^{\a}d\p_{\a}\nn.
\eea
This completes $W$ into an $s$ closed form:
\be
    W=-\frac{k}{2}e^\a L_\a.
\ee
Therefore $M=0$ in this gauge (even though $W$ is non-zero). We see that even when a quasi-invariant WZ term is used, the $D$ cocycle is still trivialized by extending the superspace appropriately. This is consistent with the general observation made in section \ref{sec:Anomalous term cohomology} that changing the WZ term does not affect the anomalous term; only changing the background will have an effect.

In the extended superspace case there are many more possibilities for the $\lambda$ gauge transformations since one can form new $\lambda$ fields using the extra coordinates. One might further extend the extended background superalgebras in this way. However, the number of possibilities for $\lambda$ is considerable and the algebras obtained can be large. Since there is currently no direct physical application of such algebras (unlike the extensions of \textit{standard} superspace considered in this paper) we will not pursue this possibility here.

\section{Comments}
\label{sec:Conclusion}

In section \ref{sec:standard superspace action}, the spectrum of algebras for the superstring was shown to contain three known extended algebras. Two of these (the Green algebra and extended Green algebra) have already found application in allowing a manifestly left invariant string WZ term to be constructed. It turns out that the entire spectrum (\ref{6:gauge fixed algebra}) of algebras for the superstring can be used this way. We find that the general solution for the WZ form $B$ takes the same form as in (\ref{2 term LI WZ form}) (the calculation is quite simple and will not be given here). In the general $p$-brane case it is possible that the cocycle approach may generate those superalgebras which allow the construction of left invariant WZ forms. Work on this issue is currently in preparation.

In this work our attention has been restricted to $p$-branes only for brevity. Similar principles to those of the $p$-brane WZ term also apply to the WZ terms of D-branes and M-branes \cite{chrys99}. The additional feature of these branes is the presence of worldvolume gauge fields. With minimal modifications to allow for these fields, the cocycle construction can also be applied to these branes. For the traditional (bosonic topology only) approach to Noether charge algebras of D-branes and M-branes see \cite{sorokin97,hammer97,Hackett03-M-brane-charges}. Work on D-brane charge algebras using the methods of this paper is currently in preparation.

The previously derived structure of the anomalous term arises in the cocycle construction in a particular choice of gauge. This simplified structure relates to the PBRS construction, where the modified algebra is written in the form of a projector \cite{townsend97}. This represents the physical situation in supergravity field theory where half the supersymmetries are broken. The work of this paper shows that allowing for $\lambda$ gauge freedom results in an expanded definition of the anomalous term. It would be interesting to revisit the PBRS construction to determine whether the new possibilities for the anomalous term can be incorporated. Ideally one would like to find a generalization of PBRS which is $\lambda$ covariant.

\subsection{Acknowledgments}

I would like to thank I. N. McArthur for helpful suggestions and critical reading of the manuscript. I would also like to thank S. M. Kuzenko for critical reading of the manuscript.

\appendix
\section{Appendices}

\subsection{Standard vielbein components}
\label{sec:app:standard vielbein components}
\subsubsection{$L_{M}{}^{A}$ components}
    $\begin{array}{ll}
    L_{m}{}^{a}=\d_{m}{}^{a}, & L_{m}{}^{\a}=0\\
    L_{\m}{}^{a}=-\half \gtu{a}{\m}, & L_{\m}{}^{\a}=\d_{\m}{}^{\a}
    \end{array}$
\subsubsection{$L_{A}{}^{ M}$ components}
    $\begin{array}{ll}
    L_{a}{}^{m}=\d_{a}{}^{m}, & L_{a}{}^{\m}=0\\
    L_{\a}{}^{m}=\half \gtu{m}{\a}, & L_{\a}{}^{\m}=\d_{\a}{}^{\m}
    \end{array}$
\subsubsection{$R_{M}{}^{A}$ components}
    $\begin{array}{ll}
    R_{m}{}^{a}=\d_{m}{}^{a}, & R_{m}{}^{\a}=0\\
    R_{\m}{}^{a}=\half \gtu{a}{\m}, & R_{\m}{}^{\a}=\d_{\m}{}^{\a}
    \end{array}$
\subsubsection{$R_{A}{}^{M}$ components}
    $\begin{array}{ll}
    R_{a}{}^{m}=\d_{a}{}^{m}, & R_{a}{}^{\m}=0\\
    R_{\a}{}^{m}=-\half \gtu{m}{\a}, & R_{\a}{}^{\m}=\d_{\a}{}^{\m}
    \end{array}$

\subsection{Green algebra vielbein components}
\label{sec:app:green algebra vielbein components}
\subsubsection{$L_{M}{}^{A}$ components}
    $\begin{array}{lll}
    L_{m}{}^{a}=\d_{m}{}^{a}, & L_{m}{}^{\a}=0, & L_{m\a}=-\gtl{m}{\a}\\
    L_{\m}{}^{a}=-\half \gtu{a}{\m}, & L_{\m}{}^{\a}=\d_{\m}{}^{\a}, & L_{\m\a}=\sixth \gtu{b}{\m}\gtl{b}{\a}\\
    L^{\m a}=0, & L^{\m\a}=0, & L^{\m}{}_{\a}=\d^{\m}{}_{\a}
    \end{array}$
\subsubsection{$L_{A}{}^{ M}$ components}
    $\begin{array}{lll}
    L_{a}{}^{m}=\d_{a}{}^{m}, & L_{a}{}^{\m}=0, & L_{a\m}=\gtl{a}{\m}\\
    L_{\a}{}^{m}=\half \gtu{m}{\a}, & L_{\a}{}^{\m}=\d_{\a}{}^{\m}, & L_{\a\m}=\third \gtu{b}{\a}\gtl{b}{\m}\\
    L^{\a m}=0, & L^{\a\m}=0, & L^{\a}{}_{\m}=\d^{\a}{}_{\m}
    \end{array}$
\subsubsection{$R_{M}{}^{A}$ components}
    $\begin{array}{lll}
    R_{m}{}^{a}=\d_{m}{}^{a}, & R_{m}{}^{\a}=0, & R_{m\a}=0\\
    R_{\m}{}^{a}=\half \gtu{a}{\m}, & R_{\m}{}^{\a}=\d_{\m}{}^{\a}, & R_{\m\a}=-x^{b}\G_{b\m\a}+\sixth    \gtu{b}{\m}\gtl{b}{\a}\\
    R^{\m a}=0, & R^{\m\a}=0, & R^{\m}{}_{\a}=\d^{\m}{}_{\a}
    \end{array}$
\subsubsection{$R_{A}{}^{M}$ components}
    $\begin{array}{lll}
    R_{a}{}^{m}=\d_{a}{}^{m}, & R_{a}{}^{\m}=0, & R_{a\m}=0\\
    R_{\a}{}^{m}=-\half \gtu{m}{\a}, & R_{\a}{}^{\m}=\d_{\a}{}^{\m}, & R_{\a\m}=x^{b}\G_{b\a\m}-\sixth    \gtu{b}{\a}\gtl{b}{\m}\\
    R^{\a m}=0, & R^{\a\m}=0, & R^{\a}{}_{\m}=\d^{\a}{}_{\m}
    \end{array}$

\subsection{Extended Green algebra vielbein components}
\label{sec:app:ext green algebra vielbein components}
\subsubsection{$L_{M}{}^{A}$ components}
    $\begin{array}{llll}
    L_{m}{}^{a}=\d_{m}^{a}, & L_{m}{}^{\a}=0, & L_{m}{}_a=0, & L_{m\a}=-\gtl{m}{\a}\\
    L_{\m}{}^{a}=-\half \gtu{a}{\m}, & L_{\m}{}^{\a}=\d_{\m}{}^{\a}, & L_{\m}{}_a=-\half \gtl{a}{\m},
        & L_{\m\a}=\third \gtu{b}{\m}\gtl{b}{\a}\\
    L^m{}^{a}=0, & L^m{}^{\a}=0, & L^m{}_a=\d^m{}_a, & L^m{}_{\a}=-\gtu{m}{\a}\\
    L^{\m a}=0, & L^{\m\a}=0, & L^{\m}{}_a=0, & L^{\m}{}_{\a}=\d^{\m}{}_{\a}
    \end{array}$
\subsubsection{$L_{A}{}^{M}$ components}
    $\begin{array}{llll}
    L_{a}{}^{m}=\d_{a}{}^{m}, & L_{a}{}^{\m}=0, & L_{am}=0, & L_{a\m}=\gtl{a}{\m}\\
    L_{\a}{}^{m}=\half \gtu{m}{\a}, & L_{\a}{}^{\m}=\d_{\a}{}^{\m}, & L_{\a m}=\half \gtl{m}{\a},
        & L_{\a\m}=\frac{2}{3} \gtu{b}{\a}\gtl{b}{\m}\\
    L^{am}=0, & L^{a\m}=0, & L^a{}_m=\d^a{}_m, & L^a{}_{\m}=\gtu{a}{\m}\\
    L^{\a m}=0, & L^{\a\m}=0, & L^{\a}{}_m=0, & L^{\a}{}_{\m}=\d^{\a}{}_{\m}
    \end{array}$
\subsubsection{$R_{M}{}^{A}$ components}
    $\begin{array}{llll}
    R_{m}{}^{a}=\d_{m}{}^{a}, & R_{m}{}^{\a}=0, & R_{m}{}_a=0, & R_{m\a}=0\\
    R_{\m}{}^{a}=\half \gtu{a}{\m}, & R_{\m}{}^{\a}=\d_{\m}{}^{\a}, & R_{\m}{}_a=\half \gtl{a}{\m},
        & R_{\m\a}=-x^{b}\G_{b\m\a}-y_b\G^b{}_{\m\a}\\&&&+\third \gtu{b}{\m}\gtl{b}{\a}\\
    R^m{}^{a}=0, & R^m{}^{\a}=0, & R^m{}_a=\d^m{}_a, & R^m{}_\a=0\\
    R^{\m a}=0, & R^{\m\a}=0, & R^{\m}{}_a=0, & R^{\m}{}_{\a}=\d^{\m}{}_{\a}
    \end{array}$
\subsubsection{$R_{A}{}^{M}$ components}
    $\begin{array}{llll}
    R_{a}{}^{m}=\d_{a}{}^{m}, & R_{a}{}^{\m}=0, & R_{a}{}_m=0, & R_{a\m}=0\\
    R_{\a}{}^{m}=-\half \gtu{m}{\a}, & R_{\a}{}^{\m}=\d_{\a}{}^{\m}, & R_{\a m}=-\half \gtl{m}{\a},
        & R_{\a\m}=+x^{b}\G_{b\a\m}+y_b\G^b{}_{\a\m}\\&&&-\third \gtu{b}{\a}\gtl{b}{\m}\\
    R^a{}^{m}=0, & R^a{}^{\m}=0, & R^a{}_m=\d^a{}_m, & R^a{}_\m=0\\
    R^{\a m}=0, & R^{\a\m}=0, & R^{\a}{}_m=0, & R^{\a}{}_{\m}=\d^{\a}{}_{\m}
    \end{array}$

\subsection{Constraints for the $p$-brane action}
\label{sec:app:algebra conditions}

Here we show that the constraints in the presence of the WZ term take the simple form (\ref{6:fermionic mod}) in both standard and extended backgrounds. The structure of the definitions of momenta may be written as the vanishing of functions $\widetilde C_{M}$ (one for each coordinate):
\be
    \widetilde C_{M}=P_{M}-P^{(NG)}_{M}-\bar B_{M},
\ee
where the terms $P^{(NG)}_{M}$ are the functions of $(Z,\dot Z)$ obtained as momenta from the NG action. However, $P^{(NG)}_{M}$ are nonzero only for the standard superspace coordinates, and they are related by:
\bea
	\label{vanishing constraint comb}
    P^{(NG)}_{\m}&=&-\half (\G^{m}\t)_{\m}P^{(NG)}_{m}\\
    \Rightarrow L_{\a}{}^{M}P^{(NG)}_{M}&=&0\nn.
\eea
For $M\neq m$, the $\widetilde C_{M}$ are constraints. Consider then the linear combinations:
\be
    \label{3:constraint linear comb}
    L_{A}{}^{M}\widetilde C_{M},\qquad M\neq m.
\ee
One can generate new sets of constraints by taking such linear combinations as long as the constraint surface so defined remains unchanged. This will be true provided that we maintain a ``linearly independent" combination of the original constraints (which are all independent in the sense of intersecting surfaces). The linear combinations (\ref{3:constraint linear comb}) will then be constraints of the form:
\be
	\widetilde C_{A}=L_{A}{}^{M}(P_{M}-\bar B_{M}),\qquad A\neq a
\ee
provided that:
\be
    \label{3:constraint match condition}
    L_{A}{}^{M}P^{(NG)}_{M}=0.
\ee
Denote the extra generators of the superalgebra by $T_{\check A}$. These generators are assumed to form an ideal. It follows that the standard coordinates do not transform under the left/right group actions generated by $T_{\check A}$. From this it follows that the components of the inverse vielbeins satisfy:
\bea
	L_{\check A}{}^{m}&=&L_{\check A}{}^{\m}=0\\
	R_{\check A}{}^{m}&=&R_{\check A}{}^{\m}=0\nn.
\eea
Using this and (\ref{vanishing constraint comb}), the required result (\ref{3:constraint match condition}) follows. The constraints can therefore be written in the form of equation (\ref{6:fermionic mod}).

\bibliographystyle{hieeetr}
\bibliography{double_complex}
\end{document}